\documentclass{PoS}

\title{Automated one-loop calculations with GoSam}

\ShortTitle{}

\author{G.~Cullen\\
 Deutsches Elektronen-Synchrotron DESY, Zeuthen, Germany\\
 E-mail: \email{gavin.cullen@desy.de}}

\author{N.~Greiner, \speaker{G.~Heinrich},  T.~Reiter\\
 Max-Planck-Institute for Physics,  Munich, Germany\\
 E-mail: \email{\{greiner,gudrun,reiterth\}@mpp.mpg.de}}

\author{G.~Luisoni\\
Institute for Particle Physics Phenomenology, University of Durham, UK\\
 E-mail: \email{gionata.luisoni@durham.ac.uk}}

\author{P.~Mastrolia\\
Max-Planck Institute for Physics, Munich, Germany;\\
Dipartimento di Fisica, Universita di Padova, Italy\\
 E-mail: \email{ppaolo@mpp.mpg.de}}

\author{G.~Ossola\\
 New York City College of Technology, City University of New York\\
 E-mail: \email{gio.ossola@gmail.com}}
 
\author{F.~Tramontano\\
 CERN, Geneva, Switzerland\\
 E-mail: \email{francesco.tramontano@cern.ch}}

\abstract{
In this talk, the program package \GOSAM{} is presented which can be used for the automated calculation of 
one-loop amplitudes for multi-particle processes.
The integrands are generated in terms of Feynman diagrams and 
can be reduced by  d-dimensional integrand-level decomposition,
or tensor reduction, or a combination of both. 
Through various examples we show that \GOSAM{} can  produce one-loop amplitudes for 
both QCD and electroweak theory; model files for theories Beyond the Standard Model 
can be linked as well.
}

\FullConference{ 10th International Symposium on Radiative Corrections 
(Applications of Quantum Field Theory to Phenomenology) - Radcor2011\\
September 26-30, 2011\\
Mamallapuram, India}

\usepackage{amsmath,amssymb,amsfonts}
\usepackage{subfigure}
\usepackage{listings}

\newcommand{\GOLEM}{{\textsc{Go\-Sam}}}
\newcommand{\GOSAM}{{\textsc{Go\-Sam}}}
\newcommand{\GOLEMVC}{{\texttt{go\-lem95C}}}
\newcommand{\QGRAF}{{\texttt{QGRAF}}}
\newcommand{\FORM}{{\texttt{FORM}}}

\newcommand{\HAGGIES}{{\texttt{hag\-gies}}}
\newcommand{\SAMURAI}{{\textsc{Sa\-mu\-rai}}}
\newcommand{\PYTHON}{{\texttt{Py\-thon}}}
\newcommand{\bea}{\begin{eqnarray}}
\newcommand{\eea}{\end{eqnarray}\noindent}
\newcommand{\nn}{\nonumber}
\newcommand{\bcen}{\begin{center}}
\newcommand{\ecen}{\end{center}}
\def\eps{\epsilon}
\def\url#1{\texttt{#1}}

\begin{document}

\section{Introduction}

Recently we have seen tremendous progress in the automation of NLO multi-leg 
calculations\,
\cite{Cullen:2011ac,Agrawal:2011tm,Cascioli:2011va,Bevilacqua:2011xh,Hirschi:2011pa,vanHameren:2009dr,Berger:2008sj}. 
In addition,   public NLO tools containing
a collection of hard-coded individual processes, like  e.g. 
MCFM\,\cite{Campbell:1999ah,Campbell:2011bn} and 
VBFNLO\,\cite{Arnold:2008rz,Arnold:2011wj,Campanario:2011cs}
have been developed and constantly enriched. 
The matching of partonic NLO predictions to parton shower Monte Carlo programs 
has also seen major advances, see e.g.  the developments in Sherpa\,\cite{Hoche:2010pf,Hoeche:2011fd},
MC@NLO/Herwig++\,\cite{Frixione:2010wd,Frixione:2010ra}, 
POWHEG-\hspace{0pt}Box \cite{Alioli:2010xd,Alioli:2011as}, 
 POWHEL\,\cite{Kardos:2011qa,Garzelli:2011vp,Kardos:2011na}.
Therefore we are progressing well  towards the aim that basically any process which may 
turn out to be important for 
the comparison of LHC findings to theory can be evaluated at NLO accuracy.

In this talk, we present the program package \GOSAM{}\,\cite{Cullen:2011ac} 
which allows the automated generation and evaluation of 
one-loop amplitudes for multi-particle processes. 
To produce results for a certain process specified by the user, 
there is an ``input card" to be edited specifying the details of the process.
Then the user can launch the generation of the source code and its compilation, 
without having to worry about internal details of the 
code. The individual program tasks are steered via python scripts.

The integrands of the one-loop amplitudes are generated in terms of Feynman diagrams, 
thus allowing to perform symbolic manipulations of the corresponding  
algebraic expressions prior to any numerical step.
For the reduction, the program offers the possibility to use either 
a $d$-dimensional extension of
the OPP method~\cite{Ossola:2006us,Ossola:2007bb,Ellis:2008ir}, 
as implemented in \SAMURAI~\cite{Mastrolia:2010nb}, 
or tensor reduction as implemented in
\GOLEMVC~\cite{Binoth:2008uq,Cullen:2011kv} interfaced 
through tensorial reconstruction at the integrand level~\cite{Heinrich:2010ax}.

The program can be used to calculate one-loop corrections within both QCD and electroweak theory. 
Beyond the Standard Model theories can be interfaced using
FeynRules~\cite{Degrande:2011ua} or \texttt{LanHEP}~\cite{Semenov:2010qt}.
The  Binoth Les Houches interface\,\cite{Binoth:2010xt} to programs providing the real radiation 
contributions is also included and has been tested in various examples. 


\section{Description of the program}

\subsection{Generation of the source code for the amplitude}

For the diagram generation 
we use the program \QGRAF~\cite{Nogueira:1991ex},  
supplemented by  \PYTHON{} routines to further 
analyse or filter the diagrams,
which  allows for example to drop diagrams whose colour factor turns out to be zero, 
or to determine the signs for diagrams with Majorana fermions.
The information about the model
is either read from the built-in Standard Model file or
is generated from a user defined \texttt{LanHEP}~\cite{Semenov:2010qt}
or Universal FeynRules Output (\texttt{UFO})~\cite{Degrande:2011ua} file.
The program also produces a  \LaTeX{} file which contains graphical
representations of all diagrams together with a summary of conventions, 
e.g. the helicity and colour basis.


The amplitude is generated in terms of algebraic expressions based on 
Feynman diagrams and then processed with a \FORM{} program, using {\tt spinney}~\cite{Cullen:2010jv}
for the spinor algebra. 
In \GOLEM{} we have implemented the 't~Hooft-\hspace{0pt}Veltman
scheme (HV) and
dimensional reduction~(DRED). In both schemes all external vectors
(momenta and polarisation vectors) are kept in four dimensions, while 
internal vectors are kept in the $d$-dimensional vector space ($d=4-2\eps$).
 For the loop momentum $q$ we introduce
the symbol $\mu^2=-\tilde{q}^2$, such that
\begin{equation}
q^2=\hat{q}^2+\tilde{q}^2=\hat{q}^2-\mu^2\;,
\end{equation}
where $\hat{q}$  lives in 4 dimensions and  the $(D-4)$-dimensional orthogonal projection
is denoted by~$\tilde{q}$.

To prepare the numerator functions of the one-loop diagrams for their numerical evaluation,
we separate the symbol $\mu^2$ and dot products involving the
momentum $\hat{q}$ from all other factors. All subexpressions which do
not depend on either $\hat{q}$ or $\mu^2$ are substituted by abbreviations,
which are evaluated only once per phase space point.
Each of the two parts is then processed by \HAGGIES~\cite{Reiter:2009ts},
which generates optimised \texttt{Fortran95} code for the numerical evaluation.
For each diagram we generate an interface to
\SAMURAI~\cite{Mastrolia:2010nb}, \GOLEMVC~\cite{Cullen:2011kv} and/or
\texttt{PJFRY}~\cite{Fleischer:2010sq,pjfry}. Our 
standard choice for the reduction is to use \SAMURAI{}~\cite{Mastrolia:2010nb},  
which usually provides a fast and stable reduction of the amplitude
to a set of coefficients of basis integrals in most of the phase space.
Furthermore, \SAMURAI{}  monitors the quality
of the reconstruction of the numerator. 
In \GOLEM{} we use this
information to trigger an alternative reduction with either
\GOLEMVC~\cite{Cullen:2011kv} or \texttt{PJFRY}~\cite{Fleischer:2010sq} whenever
these reconstruction tests fail. 
This combination of on-shell techniques and traditional tensor reduction
is achieved using tensorial reconstruction at the integrand
level~\cite{Heinrich:2010ax}. 
The tensorial reconstruction not only can cure numerical instabilities, but in some cases also
can reduce the computational cost of the reduction.
Since the reconstructed numerator is typically of a form where 
kinematics and loop momentum dependence are already separated, 
the use of a reconstructed numerator tends to be faster than the original procedure, 
in particular in cases with a large number of legs and low rank.

\subsection{Rational terms }

Terms containing the symbols $\mu^2$ or $\varepsilon$ 
in the numerator of the integrands 
can lead to a so-called
$R_2$ term~\cite{Ossola:2008xq}, which contributes to the rational part 
of the amplitude.
As we start from Feynman diagrams, we generate the $R_2$ part along with all other
contributions without the need to seperate the different parts.
In addition, we provide two different ways to calculate $R_2$, 
an \emph{implicit} and
an \emph{explicit} construction, using the fact that 
there are two ways of splitting the numerator function:
\bea
\label{eq:r2:a}
\mathcal{N}(\hat{q},\mu^2,\varepsilon)&=&
\mathcal{N}_0(\hat{q},\mu^2)+
\varepsilon\mathcal{N}_1(\hat{q},\mu^2)+\varepsilon^2\mathcal{N}_2(\hat{q},\mu^2)\\
&&\mbox{or, alternatively,}\nn\\
\label{eq:r2:b}
\mathcal{N}(\hat{q},\mu^2,\varepsilon)&=&
\hat{\mathcal{N}}(\hat{q})+
\tilde{\mathcal{N}}(\hat{q},\mu^2,\varepsilon).
\eea
The implicit
construction uses the splitting of Eq.~(\ref{eq:r2:a}) and treats
all three numerator functions~$\mathcal{N}_i$ on equal grounds.
Each of the three terms is reduced seperately in a numerical reduction
and the Laurent series of the three results are added up taking into
account the powers of $\varepsilon$. 

The explicit construction of $R_2$
is based on the assumption that each term in
$\tilde{\mathcal{N}}$ in Eq.~(\ref{eq:r2:b}) contains at least one
power of $\mu^2$ or $\varepsilon$. The expressions for those integrals
are relatively simple and known explicitly. Hence, the part of the amplitude
which originates from $\tilde{\mathcal{N}}$ is computed analytically whereas
the purely four-\hspace{0pt}dimensional part $\hat{\mathcal{N}}$ is passed to the numerical
reduction.

In the user input card, 
possible options for $R_2$ are \texttt{r2=implicit},\texttt{explicit}, \texttt{off} 
and \texttt{only}.
Using \texttt{r2=only} discards everything but
the $R_2$ term 
and puts \GOLEM{} in the position of providing $R_2$ terms
to supplement other codes which work entirely in four dimensions, 
provided they use the same gauge.

\subsection{Conventions}

To be specific, we consider the
case where the user wants to compute QCD corrections. 
In the case of electroweak corrections, 
the analogous conventions apply except that the strong coupling $g_s$ is replaced by $e$.
In the QCD case, the tree-level
matrix element squared can be written as
\begin{equation}\label{eq:amp0:def}
\vert\mathcal{M}\vert_{\text{tree}}^2=\mathcal{A}_0^\dagger\mathcal{A}_0=
(g_s)^{2b}\cdot a_0\;, 
\end{equation}
where $b=0$ is also possible.
The matrix element at one-loop level, i.e. the
interference term between tree-level and one-loop, can be written as
\begin{multline}\label{eq:amp1:def}
\vert\mathcal{M}\vert_{\text{1-loop}}^2=
\mathcal{A}_1^\dagger\mathcal{A}_0+
\mathcal{A}_0^\dagger\mathcal{A}_1=
2\cdot\Re(\mathcal{A}_0^\dagger\mathcal{A}_1)=
\frac{\alpha_s(\mu)}{2\pi}\frac{(4\pi)^\varepsilon}{\Gamma(1-\varepsilon)}
\, (g_s)^{2b}\,\left[%
c_0+\frac{c_{-1}}{\varepsilon}+\frac{c_{-2}}{\varepsilon^2}
+\mathcal{O}(\varepsilon)%
\right]\;.
\end{multline}
A call to the subroutine \texttt{samplitude} returns an array
consisting of the four numbers $(a_0, c_0, c_{-1}, c_{-2})$
in this order. 
The average over initial state colours and helicities is included 
in the default setup. Renormalisation is included depending on the options 
chosen by the user, for a more detailed description we refer to \cite{Cullen:2011ac}.

\section{Installation and Usage}

\subsection{Installation}
The user can download the code \GOLEM{} either as a tar-ball
    or from the subversion repository at
    \url{http://projects.hepforge.org/gosam/}\, .
    The build process and in\-stal\-la\-tion of \GOLEM{} is controlled by
    \PYTHON{} \texttt{Dist\-utils}, while the build process for the libraries 
    \SAMURAI{} and \GOLEMVC{}
    is controlled by \texttt{Autotools}.
To install \GOLEM{}, the user needs to run\\
{\tt python setup.py install --prefix MYPATH}.\\
For more details we direct the user to
the \GOLEM{} reference manual coming with the code.\\
On top of a standard Linux environment, the programs
\FORM~\cite{Vermaseren:2000nd}, 
version~$\geq3.3$, and
\QGRAF~\cite{Nogueira:1991ex} need to be installed on the system.
Further, at least one of the libraries
\SAMURAI{}~\cite{Mastrolia:2010nb} and \GOLEMVC{}~\cite{Cullen:2011kv}
needs to be present at compile time of the generated code.    
For the user's convenience we have prepared a package containing
    \SAMURAI{} and \GOLEMVC{} together with the integral libraries
    \texttt{One\-LOop}\,\cite{vanHameren:2010cp},
    \texttt{QCD\-Loop}\,\cite{Ellis:2007qk} and
    \texttt{FF}\,\cite{vanOldenborgh:1989wn}.
   The package \texttt{gosam-\hspace{0pt}contrib-\hspace{0pt}1.0.tar.gz}
   containing all these libraries is available
   for download from\\
    \url{http://projects.hepforge.org/gosam/}.

\subsection{Usage}

In order to generate the code for a process,
the user needs to prepare an input file, 
called \textit{process card} in the following, which contains
\begin{itemize}
\item process specific information, such as a list of initial and
      final state particles, their helicities (optional) 
      and the order of the coupling constants;
\item scheme specific information and approximations, such as
      the regularisation and renormalisation schemes,
      the underlying model,
      masses and widths which are set to zero, 
      the selection of subsets of diagrams, etc; 
\item system specific information, such as paths to programs and libraries
      or compiler options;
\item optional information for optimisations which control the code generation.
\end{itemize}

If the process card is called \texttt{gosam.in}, it can be invoked by 
the command \texttt{gosam.py gosam.in}. 
All further steps are controlled by the generated make files;
in order to generate and compile all files relevant for the matrix element
one needs to invoke
\begin{lstlisting}
make compile
\end{lstlisting}
The generated code can be tested with the program \texttt{matrix/test.f90}.
The following sequence of commands will compile and run the test program:
\begin{lstlisting}
cd matrix
make test.exe
./test.exe
\end{lstlisting}
The numbers printed by the test program are, 
in this order, $a_0$, $c_0/a_0$, $c_{-1}/a_0$,
$c_{-2}/a_0$ and the pole parts calculated from the infrared insertion
operator~\cite{Catani:1996vz,Catani:2000ef}.
One can also generate a pictorial representation of all generated diagrams using
the command
\begin{lstlisting}
make doc
\end{lstlisting}
which generates the file \texttt{doc/process.ps}

\subsection{Interfacing the code}
\subsubsection{Using the BLHA Interface}

The so-called \emph{Binoth Les Houches Accord} (BLHA)~\cite{Binoth:2010xt}
defines an interface for a standardized communication between
one-loop programs (OLP) and Monte Carlo (MC) tools. 
\GOLEM{} can act as an OLP in the framework of the BLHA, such that 
the calculation of complete cross sections is straightforward.

In general, the MC writes an order file, called for  example 
\texttt{olp\_order.lh}, and invokes the script \texttt{gosam.py}
as follows:
\begin{lstlisting}
gosam.py --olp olp_order.lh
\end{lstlisting}
The invocation of \texttt{gosam.py} generates a set of files which
can be compiled with a generated make file. The BLHA routines
are defined in the \texttt{Fortran} module \texttt{olp\_module} but can also
be accessed from \texttt{C} programs. 

\subsubsection{Using External Model Files}

\GOLEM{} can also make use of model files generated
by either \texttt{Feynrules}~\cite{Christensen:2008py}
in the \texttt{UFO} format~\cite{Degrande:2011ua} or by
{\texttt LanHEP}~\cite{Semenov:2010qt}. 
The particles can be specified by their PDG code.
Details about how to import these model files are described in
the \GOLEM{} reference manual. 
Precompiled \texttt{MSSM\_UFO} and \texttt{MSSM\_LHEP} files 
can also be found in the subdirectory \texttt{examples/model}.
The \texttt{examples} directory also contains examples showing how
\texttt{UFO} or {\texttt LanHEP} model files are imported.

\section{Examples and validation}
The code generated by \GOLEM{} has been compared to a considerable number of processes 
available in the literature, as listed in Table \ref{Table:compare}. 
Many of these processes are also included as examples in the code, 
including reference values. 

\begin{figure}[htb]
\subfigure[Transverse momentum of the leading jet for $W^{-}$+\,jet production at LHC with $\sqrt{s}=7$~TeV.]{\label{fig:gosamsherpa:pt}%
\includegraphics[width=0.53\textwidth]{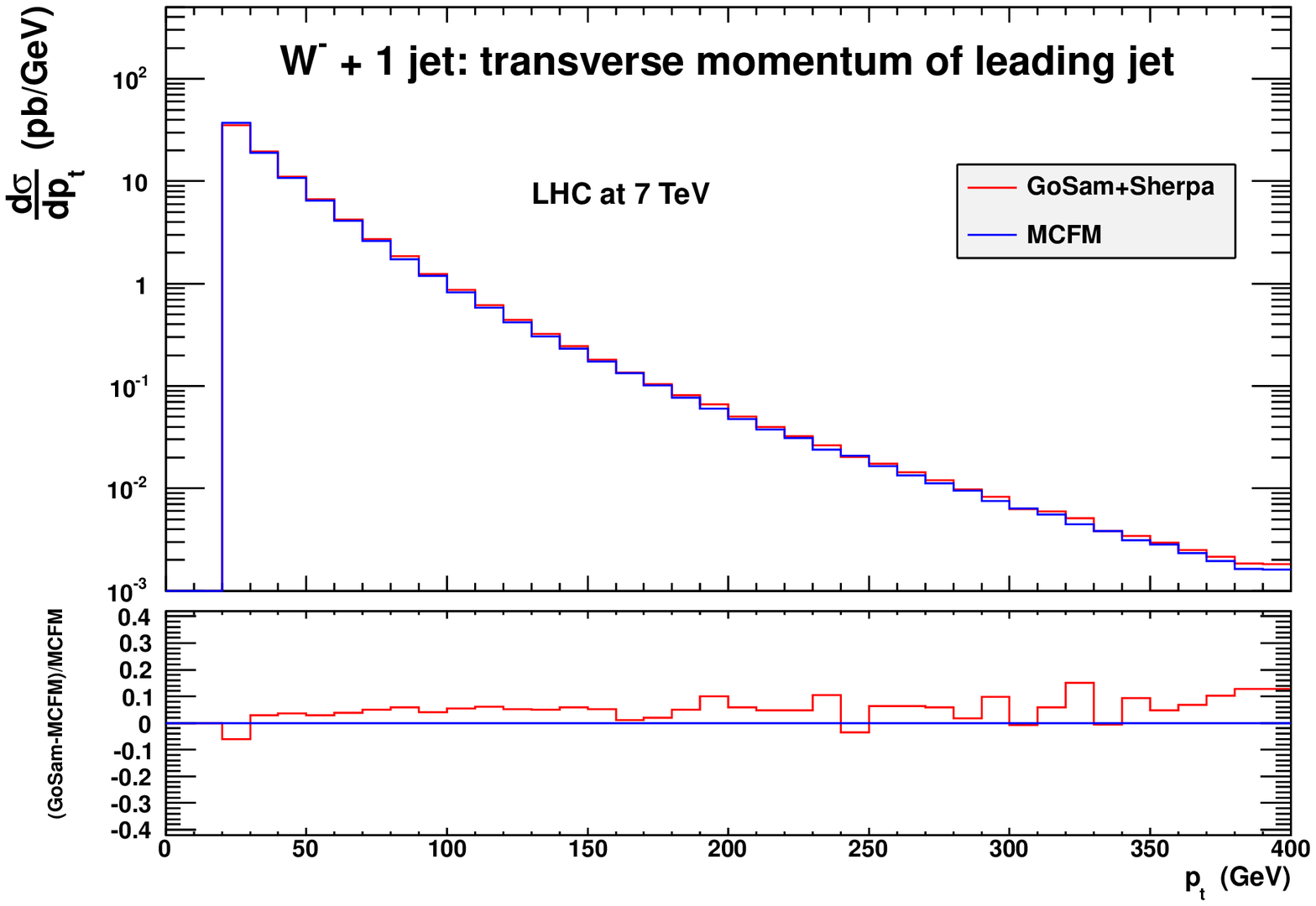}}
\subfigure[Pseudorapidity of the leading jet for $W^{-}$+\,jet production
at LHC with $\sqrt{s}=7$~TeV.]{\label{fig:gosamsherpa:eta}%
\includegraphics[width=0.53\textwidth]{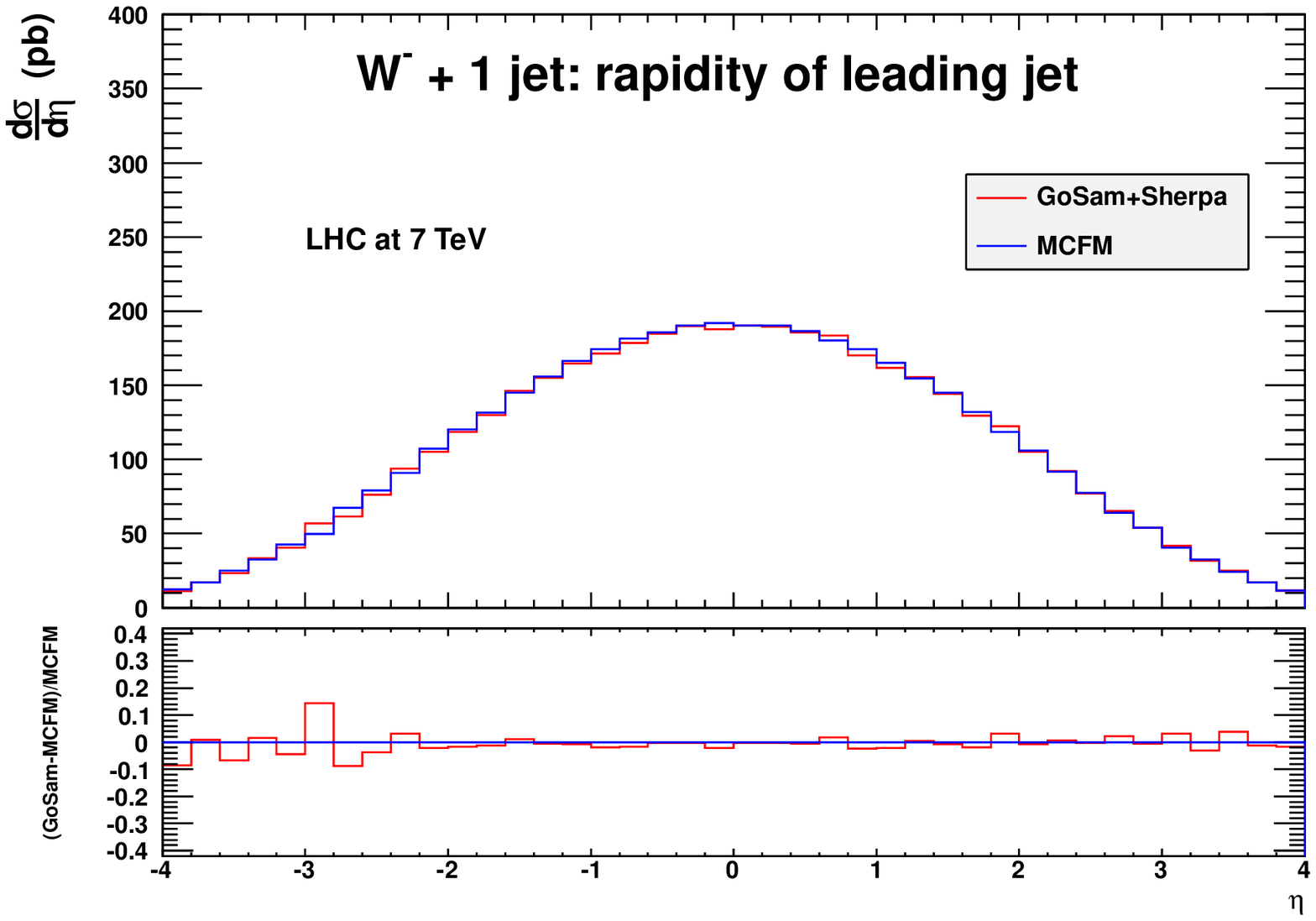}}
\caption{NLO calculation of $W^{-}+$\,jet production at LHC using \GOLEM{} 
interfaced with \texttt{SHERPA} via the Binoth Les Houches interface and compared to MCFM.}
\end{figure} 

\begin{table}
\begin{center}
\begin{tabular}{|ll|}
\hline
\bf process&\bf checked with Ref.\\
\hline
$e^+e^-\to u\overline{u}$&\cite{Ellis:1991qj}\\
$e^+e^-\to t\overline{t}$&\cite{Jersak:1981sp,Catani:2002hc}, own analytic calculation\\
$u\overline{u}\to d\overline{d}$&\cite{Ellis:1985er,Hirschi:2011pa}\\
$g g \to gg$&\cite{Binoth:2006hk}\\
$g g \to gZ$&\cite{vanderBij:1988ac}\\
$b g \to H\,b$&\cite{Campbell:2002zm,Hirschi:2011pa}\\
$\gamma \gamma \to \gamma \gamma $ (W loop) &\cite{Gounaris:1999gh}\\
$\gamma \gamma \to \gamma \gamma \gamma \gamma $ (fermion loop) &\cite{Bernicot:2008nd}\\
$pp \to t\overline{t}$&\cite{Hirschi:2011pa}, MCFM
\cite{Campbell:1999ah,Campbell:2011bn}\\
$pp\to W^\pm\,j$ (QCD corr.) &\cite{Campbell:1999ah,Campbell:2011bn} \\
$pp\to W^\pm\,j$ (EW corr.) &for IR poles: \cite{Kuhn:2007cv,Gehrmann:2010ry} \\
$pp\to W^\pm\,t$ &\cite{Campbell:1999ah,Campbell:2011bn} \\
$pp\to W^\pm\,jj$ &\cite{Campbell:1999ah,Campbell:2011bn} \\
$pp\to W^\pm b\bar{b}$ (massive b) &\cite{Campbell:1999ah,Campbell:2011bn} \\
$e^+e^-\to e^+e^-\gamma$ (QED)&\cite{Actis:2009uq}\\
$pp \to H\,t\overline{t}$&\cite{Hirschi:2011pa}\\
$pp \to Z\,t\overline{t}$&\cite{Bevilacqua:2011xh}\\
$pp\to W^+W^+jj$&\cite[v3]{Melia:2010bm}\\
$pp\to b\overline{b} b\overline{b}$ &\cite{Binoth:2009rv,Greiner:2011mp}\\
$pp\to W^+W^- b\overline{b}$ &\cite{Hirschi:2011pa,vanHameren:2009dr}\\
$pp \to t\overline{t}b\overline{b}$&\cite{Hirschi:2011pa,vanHameren:2009dr}\\
$u\overline{d} \to W^+ ggg$&\cite{vanHameren:2009dr}\\
\hline
\end{tabular}
\end{center}
\caption{Processes for which \GOLEM{} has been compared to the literature.\label{Table:compare}}
\end{table}

As an example for the  {\it Binoth Les Houches Accord} interface of \GOSAM{} we present results for 
the QCD corrections to $W^{-}+1$ jet production, obtained by linking \GOSAM{} with 
\texttt{SHERPA}~\cite{Gleisberg:2008ta}. 
Results for the transverse momentum and rapidity distribution 
of the leading jet are shown in Figs.~\ref{fig:gosamsherpa:pt} and~\ref{fig:gosamsherpa:eta}.
The comparison with MCFM\,\cite{Campbell:1999ah,Campbell:2011bn} shows perfect agreement.
Furthermore, \texttt{SHERPA} offers the possibility to match NLO calculations with a 
parton shower~\cite{Hoche:2010pf,Hoeche:2011fd}.


\section{Conclusions}

We have presented the program \GOSAM{} 
which can produce  code  for  one-loop amplitudes for multi-particle processes 
in an automated way.
The program is publicly available at\\
\url{http://projects.hepforge.org/gosam/} 
and can be used to calculate one-loop amplitudes 
within QCD, electroweak theory, or other models 
which can be imported  via an interface to
LanHEP  or FeynRules. 
 Monte Carlo programs for the real radiation can be easily linked 
 via the interface defined by the   Binoth Les Houches Accord.
 
The amplitudes are generated in terms of Feynman diagrams
and can be reduced by unitarity based reduction
at integrand level or traditional tensor reduction, 
or a combination of the two approaches. 
Further, the user can choose among different libraries for the master integrals 
and different  regularisation and  renormalisation schemes.

The calculation of the rational terms can proceed either 
together with the same numerical reduction as the rest of the amplitude, 
or  before any reduction,  
using analytic information on the integrals which can potentially give rise to 
a rational part. This feature also allows to use the code for the calculation
of rational parts only.
Moreover, the \GOSAM{} generator can  produce code for processes
which include unstable particles, i.e. intermediate states with complex masses.

 \GOSAM{} is very well suited for the automated matching of Monte Carlo programs 
 to NLO virtual amplitudes, and therefore hopefully will be used widely 
 as a module to produce 
 differential cross sections for multi-particle processes 
 which can be compared directly to experiment.

\subsection*{Acknowledgments}
G.H. would like to thank the organizers of RADCOR 2011 for their 
hospitality and the well organized conference.
G.C. and G.L. were supported by the British Science and Technology Facilities
Council (STFC).
The work of G.C. was supported by DFG
Sonderforschungsbereich Transregio 9, Computergest\"utzte Theoretische Teilchenphysik.
N.G. was supported in part by the U.S. Department of Energy under contract
No. DE-FG02-91ER40677.
P.M. and T.R. were supported by the Alexander von
Humboldt Foundation, in the framework of the Sofja Kovaleskaja Award Project
``Advanced Mathematical Methods for Particle Physics'', endowed by the German
Federal Ministry of Education and Research.
The work of G.O. was supported in part by the National Science Foundation
under Grant PHY-0855489 and PHY-1068550.
The research of F.T. is supported by Marie-Curie-IEF, project:
``SAMURAI-Apps''.
We also acknowledge the support of the Research Executive Agency (REA)
of the European Union under the Grant Agreement number
PITN-GA-2010-264564 (LHCPhenoNet).


\providecommand{\newblock}{}

\end{document}